\documentclass[twoside,11pt]{article}

\usepackage{booktabs}
\usepackage{color}
\usepackage{caption}
\usepackage{dirtytalk}
\usepackage{jmlr2e}
\usepackage[acronym,toc,shortcuts]{glossaries}
\usepackage{microtype}
\usepackage{mathtools}
\usepackage{multirow}
\usepackage{gensymb}

\newacronym{dl}{DL}{Deep Learning}
\newacronym{dp}{DP}{Differential Privacy}
\newacronym{dpsgd}{DP-SGD}{Differentially Private Stochastic Gradient Descent}
\newacronym{ecnn}{ECNN}{Equivariant CNN}

\jmlrheading{}{}{}{}{}{Florian A. Hölzl, Daniel Rueckert and Georgios Kaissis}

\ShortHeadings{Differentially Private Equivariant Deep Learning for Medical Image Analysis}{Hölzl, Rueckert and Kaissis}
\firstpageno{1}

\begin{document}

\title{Bridging the Gap: Differentially Private Equivariant Deep Learning for Medical Image Analysis}

       
\author{\name Florian A. Hölzl, Daniel Rueckert, Georgios Kaissis \\\email \{florian.hoelzl, daniel.rueckert, g.kaissis\}@tum.de \\
       \addr Institute for Artificial Intelligence in Medicine\\
       Technical University of Munich}       


\maketitle

\begin{abstract}
Machine learning with formal privacy-preserving techniques like \gls{dp} allows one to derive valuable insights from sensitive medical imaging data while promising to protect patient privacy, but it usually comes at a sharp privacy-utility trade-off.
In this work, we propose to use steerable equivariant convolutional networks for medical image analysis with \gls{dp}. Their improved feature quality and parameter efficiency yield remarkable accuracy gains, narrowing the privacy-utility gap.
\end{abstract}


\section{Introduction}
Deep learning-based medical imaging analysis has so-far been unable to fully leverage the advances in other fields of computer vision because many advanced applications require large datasets to obtain sufficient accuracy for diagnostic use.
In medicine, procuring such datasets is problematic, as patient privacy mandates minimising the amount of data collected and/or shared.
Privacy-enhancing technologies allow one to derive insights from sensitive datasets while protecting the individuals, and represent the best chance to date to incentivize data sharing in an ethical and responsible manner.
Deep learning with \glsfirst{dp} \citep{dwork2014, abadi2016}, a gold-standard technique for privacy preservation, enables analysts to train predictive models which can be shared for diagnostic purposes while offering formal guarantees about how much information can be extracted from their representations.
However, \gls{dp} leads to sharp utility trade-offs, as gradients are norm-bounded and noised, while noise scales proportionally to the number of model parameters, making training of large models disproportionately difficult. 
Thus, advanced deep learning techniques, able to learn robust and generalizable features under \gls{dp} must be developed~\citep{tramer2020}. \gls{ecnn} architectures (to Euclidean transformations) \citep{cohen2016, cohen2017} exhibit greatly increased data efficiency, improved generalization and high parameter efficiency, especially in domains with high degrees of intra-image symmetry such as medical imaging. 
So far, no works have investigated the training of \gls{ecnn}s under \gls{dp}, even though their characteristics render them highly attractive for this use-case. In this paper, we initiate an empirical investigation into \gls{dp} \gls{ecnn}s using steerable kernels. 
Our results demonstrate that \gls{ecnn}s have superior performance with a narrowed privacy-utility gap, even at smaller model sizes. We show that these beneficial attributes are a result of lower gradient sparsity, improved feature extraction characteristics and better model calibration.

\section{Results}
For all experiments, we utilized the ResNet-9 architecture and privacy parameters ($\varepsilon=7.42, \delta=10^{-5}$) of~\citet{klause2022}, adapted for equivariant training, and the MedMNIST dataset~\citep{yang2021}.
Steerable \gls{ecnn}s were implemented with regular representations and group pooling for the cyclic groups $C_N$, with $N$, the number of discrete rotations, restricted to $N/2$ after the first residual layer~\citep{weiler2019}, and with induced irreducible representations of frequency $1$, induced gated non-linearities and induced norm-pooling for the $O(2)$ group. All results are summarized in Table~\ref{tab:medmnist}.

\noindent
\textbf{Superior \gls{dp} classification performance:} Steerable \gls{ecnn}s generally outperformed their non-equivariant counterparts when training with \gls{dp}, most notably on the \textit{colon pathology} dataset, where an accuracy increase of $\approx9\%$ was observed.
The $O(2)$ group performed best on the \textit{blood cell} and \textit{dermatology} datasets, indicating that the benefit of higher equivariance outweighed the additional noise required due to the larger model size ($240k$ vs. $90k$ parameters for $C_{16}$).
Out of the cyclic groups, the $C_4$ group performed especially well, indicating a good trade-off between equivariance and model size.
Augmentations (performed as in \citet{hoffer2020}) --on average-- increased accuracy by $0.56\%$ for \gls{dp} \gls{ecnn}s but decreased performance by $0.15\%$ for non-equivariant models.

\noindent
\textbf{Reduced private/non-private accuracy gap:} The \gls{dp} validation accuracy was --on average-- $5.5\%$ lower for conventional CNNs compared to non-private training.
\gls{ecnn}s diminished this gap, with a reduction of only $2.8\%$, approaching the non-\gls{dp} baseline set by~\citet{yang2021} on MedMNIST using much larger architectures like ResNet-50.

\begin{figure}[ht]
    \includegraphics[width=.49\textwidth]{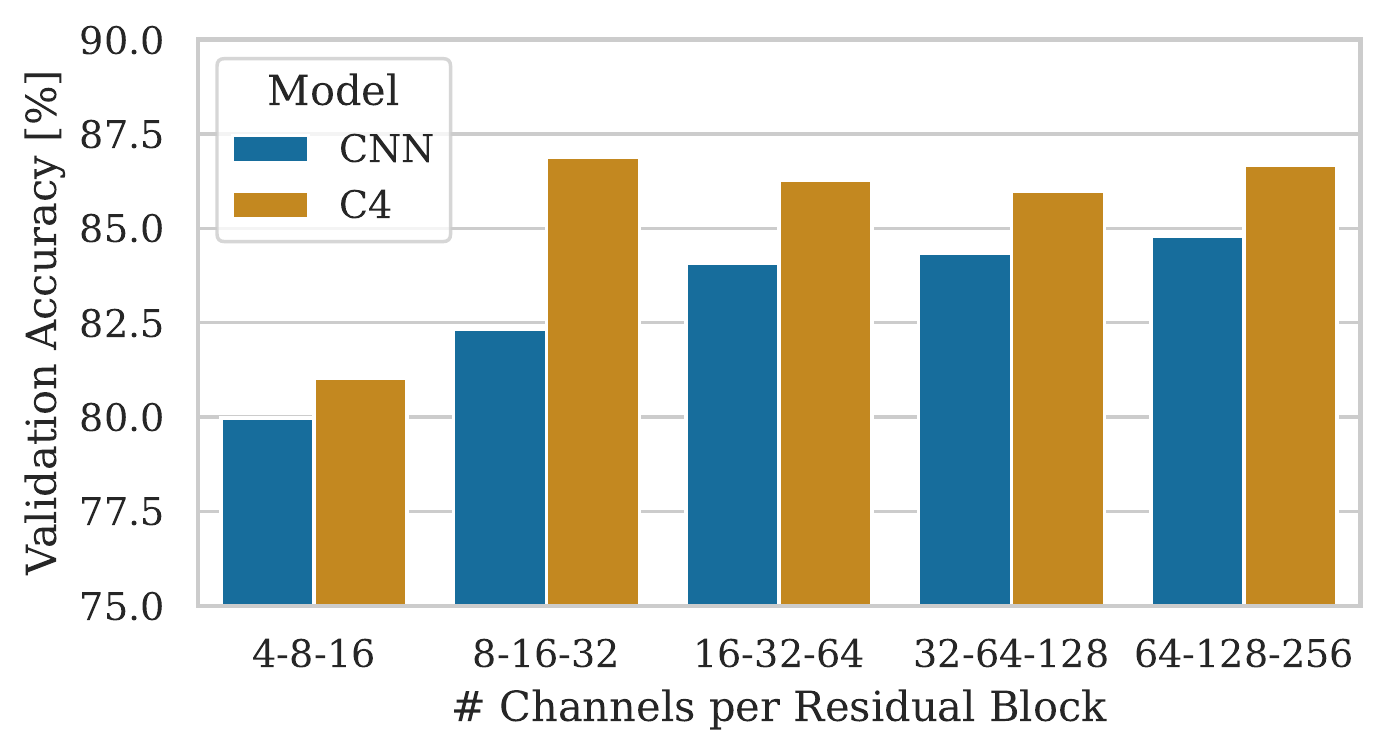}\hfill
    \includegraphics[width=.49\textwidth]{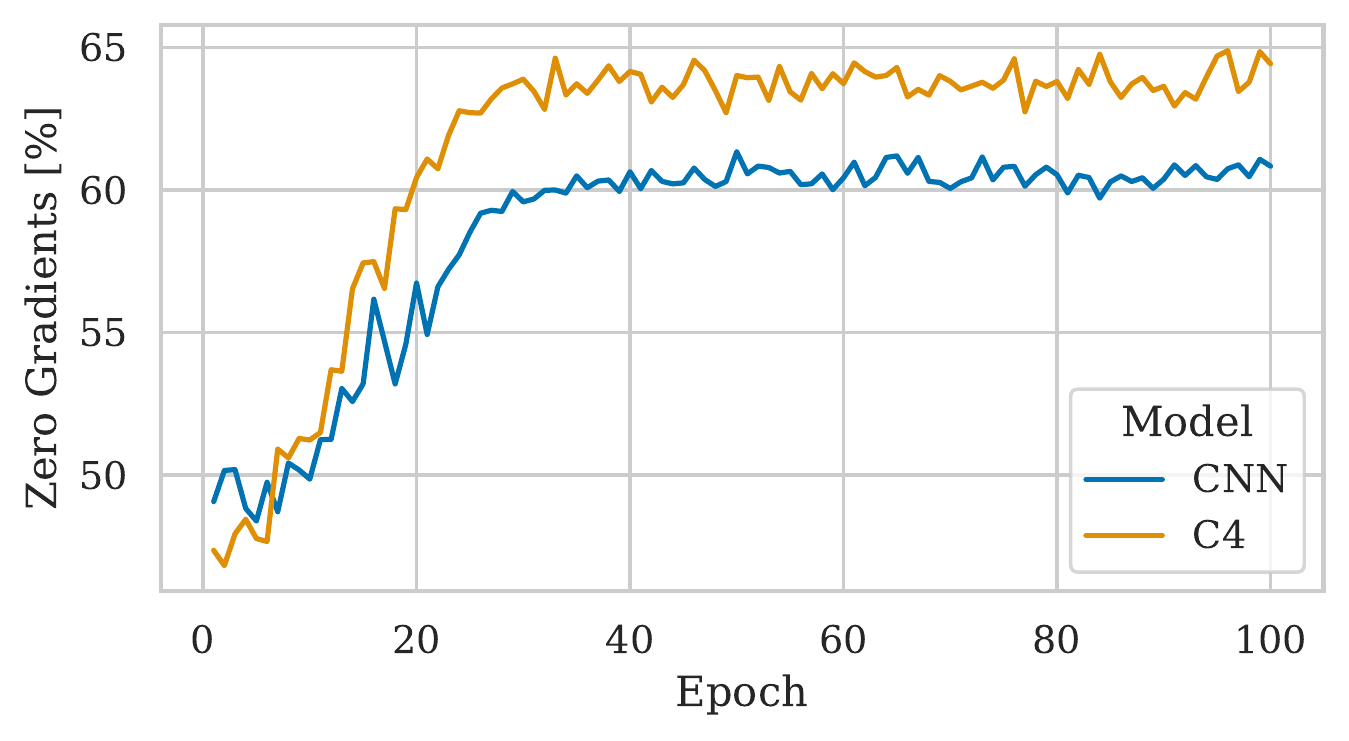}\hfill
    \caption{Results on the \textit{colon pathology} dataset for different \gls{dp} conventional and $C_4$ equivariant model widths (left). $\ell_{10^{-5}}^0$ gradient sparsity during training (right).}
    \label{fig:model_characteristics}
\end{figure}

\noindent
\textbf{Improved accuracy despite smaller model size:} \gls{dp} \gls{ecnn}s not only outperformed larger conventional CNNs, but even larger \gls{ecnn}s. For instance, the $35k$ parameter $C_{4}$ \gls{dp} \gls{ecnn} outperformed both the much larger $2,5M$ parameter conventional CNN but also the $2,3M$ parameter \gls{dp} \gls{ecnn} (Figure~\ref{fig:model_characteristics} left).

\begin{table}
\scriptsize
\centering
\begin{tabular}{lllllll}
\multirow{2}{*}{Dataset} & \multirow{2}{*}{Data Modality} & \multirow{2}{*}{Group} & \multicolumn{2}{l}{No Augmentation} & \multicolumn{2}{l}{Augmentation} \\
 &  &  & Non-DP & DP & Non-DP & DP \\ 
\toprule
& \multirow{6}{*}{Blood Cell {Microscope}} & $\{e\}$ & 94.05\% & 89.19\% & 96.52\% & 90.61\% \\
 &  & $C_{1}$ & 93.52\% & 89.98\% & 95.57\% & 90.47\% \\
 &  & $C_{4}$ & 95.35\% & 91.30\% & 96.67\% & 91.39\% \\
 &  & $C_{8}$ & 95.32\% & 86.56\% & 96.30\% & 92.39\% \\
 &  & $C_{16}$ & 95.74\% & 91.48\% & 96.16\% & 91.67\% \\
 &  & $O(2)$ & \textbf{96.13\%} & \textbf{92.51\%} & \textbf{96.71\%} & \textbf{93.73\%} \\ 
\cmidrule{2-7}
 \multirow{6}{*}{{MedMNIST}} & \multirow{6}{*}{{Dermatoscope}} & $\{e\}$ & 76.79\% & 71.78\% & \textbf{78.48\%} & 72.41\% \\
 &  & $C_{1}$ & 75.77\% & 72.84\% & 76.48\% & 72.27\% \\
 &  & $C_{4}$ & 75.18\% & 73.25\% & 77.84\% & \textbf{74.17\%} \\
 &  & $C_{8}$ & 76.48\% & 72.79\% & 77.58\% & 66.88\% \\
 &  & $C_{16}$ & 76.86\% & 72.53\% & 77.57\% & 70.64\% \\
 &  & $O(2)$ & \textbf{77.24\%} & \textbf{73.29\%} & 77.98\% & 72.45\% \\ 
\cmidrule{2-7}
 & \multirow{6}{*}{{Colon} {Pathology}} & $\{e\}$ & 83.53\% & 80.11\% & \textbf{85.75\%} & 79.41\% \\
 &  & $C_{1}$ & 82.35\% & 83.57\% & 82.22\% & 83.68\% \\
 &  & $C_{4}$ & 83.97\% & 86.89\% & 83.43\% & 83.52\% \\
 &  & $C_{8}$ & 85.11\% & 88.13\% & 85.23\% & 88.75\% \\
 &  & $C_{16}$ & \textbf{85.88\%} & \textbf{89.14\%} & 85.38\% & \textbf{88.94\%} \\
 &  & $O(2)$ & 82.12\% & 77.91\% & 80.87\% & 80.64\% \\ 
\bottomrule
\end{tabular}
\captionof{table}{Validation accuracies of the ResNet-9 for a model layout of 8-16-32 channels.}\label{tab:medmnist}
\end{table}

\begin{figure}[ht]
 \centering
    \includegraphics[width=0.65\textwidth]{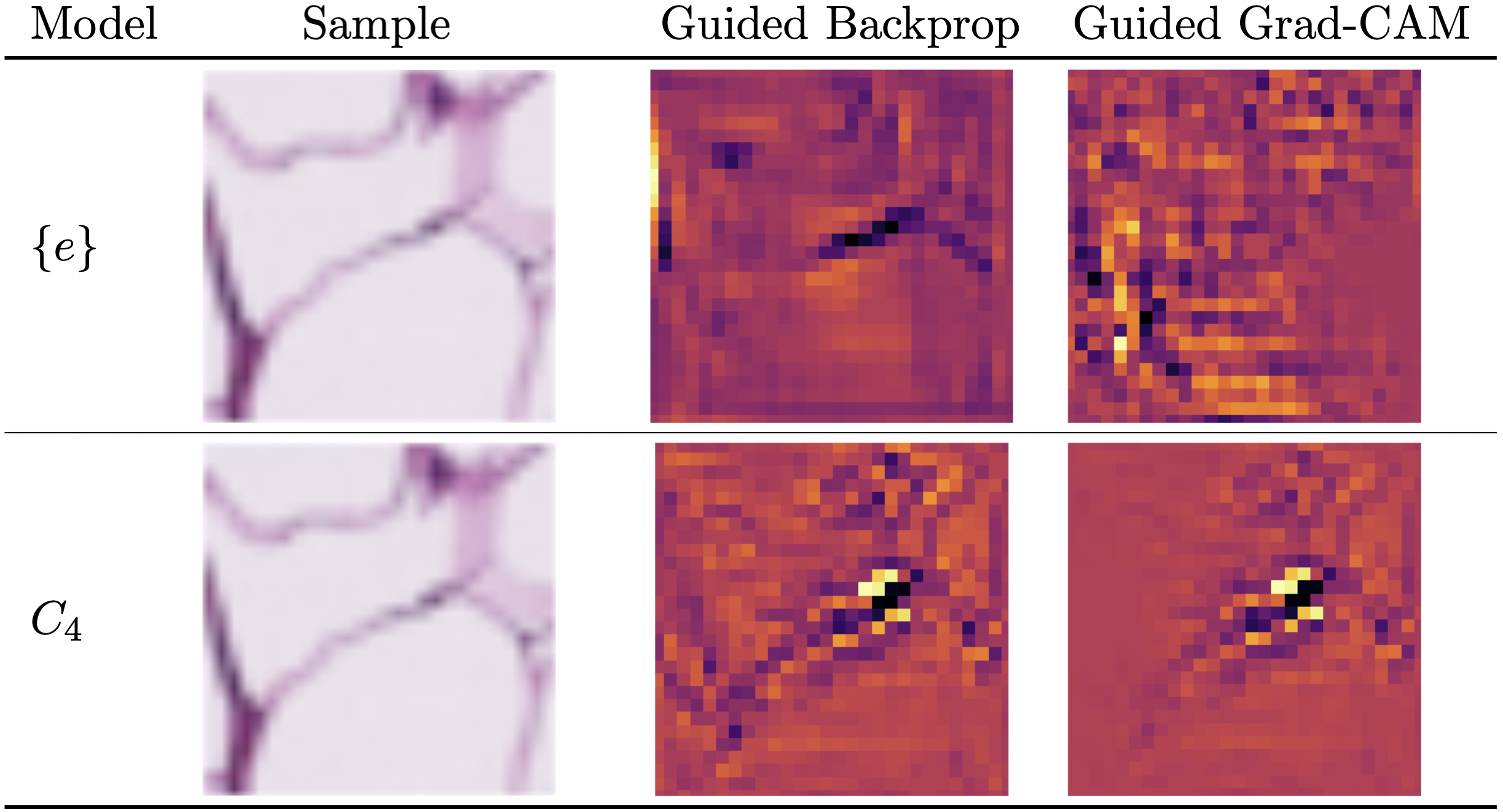}\hfill
    \caption{Guided Backprop/Grad-CAM of the final model layer. \textit{Colon pathology} dataset.}
    \label{fig:sample_results}
\end{figure}

\noindent
\textbf{Result Interpretation:} Figure~\ref{fig:model_characteristics} (right) shows, that \gls{dp} \gls{ecnn}s converge faster during training, updating fewer weights.
This is indicated by a higher gradient sparsity, as measured by the $\ell_{\epsilon}^{0}$-norm, for the \gls{ecnn} with the $C_{4}$ group.
\gls{dp} \gls{ecnn}s also exhibited better model calibration (e.g. $41.17\%$ lower Brier score in the aforementioned model). 
Moreover, a more robust feature extraction by the steerable kernels was observed, indicated by a higher magnitude filter impulse response (FIR) and an improved co-location of the FIR with the input features used for prediction (Figure~\ref{fig:sample_results}).
These findings corroborate the higher sample and feature efficiency of \gls{ecnn}s even in the \gls{dp} setting, explaining their superior performance.

\section{Conclusion}
Our preliminary findings indicate that \gls{dp} \gls{ecnn}s can help close the privacy-utility gap in medical imaging tasks through the beneficial characteristics of equivariant convolutions and the resulting feature, sample and parameter efficiency. In future work, we intend to investigate a larger variety of datasets and models in medical imaging and beyond.

\newpage




\bibliography{paper}

\begin{thebibliography}{9}
\providecommand{\natexlab}[1]{#1}
\providecommand{\url}[1]{\texttt{#1}}
\expandafter\ifx\csname urlstyle\endcsname\relax
  \providecommand{\doi}[1]{doi: #1}\else
  \providecommand{\doi}{doi: \begingroup \urlstyle{rm}\Url}\fi

\bibitem[Abadi et~al.(2016)Abadi, Chu, Goodfellow, McMahan, Mironov, Talwar,
  and Zhang]{abadi2016}
Mart{\'i}n Abadi, Andy Chu, Ian~J. Goodfellow, H.~B. McMahan, Ilya Mironov,
  Kunal Talwar, and Li~Zhang.
\newblock {Deep Learning with Differential Privacy}.
\newblock \emph{Proceedings of the 2016 ACM SIGSAC Conference on Computer and
  Communications Security}, 2016.

\bibitem[Cohen and Welling(2016)]{cohen2016}
Taco Cohen and Max Welling.
\newblock {Group Equivariant Convolutional Networks}.
\newblock In \emph{International Conference of Machine Learning}, 2016.

\bibitem[Cohen and Welling(2017)]{cohen2017}
Taco Cohen and Max Welling.
\newblock {Steerable CNNs}.
\newblock In \emph{International Conference on Learning Representations}, 2017.

\bibitem[Dwork and Roth(2014)]{dwork2014}
Cynthia Dwork and Aaron Roth.
\newblock {The Algorithmic Foundations of Differential Privacy}.
\newblock \emph{Found. Trends Theor. Comput. Sci.}, 9:\penalty0 211--407, 2014.

\bibitem[Hoffer et~al.(2020)Hoffer, Ben-Nun, Hubara, Giladi, Hoefler, and
  Soudry]{hoffer2020}
Elad Hoffer, Tal Ben-Nun, Itay Hubara, Niv Giladi, Torsten Hoefler, and Daniel
  Soudry.
\newblock {Augment Your Batch: Improving Generalization Through Instance
  Repetition}.
\newblock \emph{2020 IEEE/CVF Conference on Computer Vision and Pattern
  Recognition (CVPR)}, pages 8126--8135, 2020.

\bibitem[Klause et~al.(2022)Klause, Ziller, Rueckert, Hammernik, and
  Kaissis]{klause2022}
Helena Klause, Alexander Ziller, Daniel Rueckert, Kerstin Hammernik, and
  Georgios Kaissis.
\newblock {Differentially private training of residual networks with scale
  normalisation}.
\newblock In \emph{ICML Theory and Practice of Differential Privacy Workshop},
  2022.

\bibitem[Tram{\`{e}}r and Boneh(2020)]{tramer2020}
Florian Tram{\`{e}}r and Dan Boneh.
\newblock {Differentially Private Learning Needs Better Features (or Much More
  Data)}.
\newblock In \emph{International Conference on Learning Representations}, 2020.

\bibitem[Weiler and Cesa(2019)]{weiler2019}
Maurice Weiler and Gabriele Cesa.
\newblock {General E(2)-Equivariant Steerable CNNs}.
\newblock \emph{Advances in Neural Information Processing Systems}, 32, 2019.

\bibitem[Yang et~al.(2021)Yang, Shi, Wei, Liu, Zhao, Ke, Pfister, and
  Ni]{yang2021}
Jiancheng Yang, Rui Shi, D.~Wei, Zequan Liu, Lin Zhao, Bilian Ke, Hanspeter
  Pfister, and Bingbing Ni.
\newblock {MedMNIST v2: A Large-Scale Lightweight Benchmark for 2D and 3D
  Biomedical Image Classification}.
\newblock \emph{ArXiv}, abs/2110.14795, 2021.

\end{thebibliography}

\end{document}